\documentclass[12pt,a4paper,oneside]{article}
\usepackage[utf8]{inputenc}
\usepackage{graphicx}
\usepackage{color}
\usepackage{subscript}
\usepackage[headings]{fullpage}
\usepackage{amsmath,amssymb}
\usepackage{authblk}

\newcommand{\red}[1]{{\color{black}#1}}    

\begin{document}
\markboth{\small S. Dirkmann et al., Kinetic Simulation of Filament Growth Dynamics \dots}{\small S. Dirkmann et al., Kinetic Simulation of Filament Growth Dynamics \dots}
\title{Kinetic Simulation of Filament Growth Dynamics in Memristive Electrochemical Metallization Devices}
\author[1]{Sven Dirkmann}
\author[2]{Martin Ziegler}
\author[2]{Mirko Hansen}
\author[2]{Hermann Kohlstedt}
\author[1]{\\ Jan Trieschmann}
\author[1]{Thomas Mussenbrock\thanks{Corresponding author: thomas.mussenbrock@rub.de}}
\affil[1]{Institute of Theoretical Electrical Engineering,
Department of Electrical Engineering and Information 
Science, Ruhr University Bochum, D-44780 Bochum, Germany}
\affil[2]{Nanoelectronics Group,
Institute of Electrical and Information Engineering,
Christian Albrechts University Kiel, D-24143 Kiel, Germany}

\maketitle

\begin{abstract}
In this work we report on kinetic Monte-Carlo calculations of resistive switching and the underlying growth dynamics of filaments in an electrochemical metallization device consisting of an Ag/TiO$_2$/Pt sandwich-like thin film system. The developed model is not limited to i) fast time scale dynamics and ii) only one growth and dissolution cycle of metallic filaments. In particular, we present results from the simulation of consecutive cycles. We find that the numerical results are in excellent agreement with experimentally obtained data. Additionally, we observe an unexpected filament growth mode which is in contradiction to the widely acknowledged picture of filament growth, but consistent with recent experimental findings.
\end{abstract}

\clearpage

\section{Introduction}

It has been shown more than 50 years ago that under certain experimental conditions metal oxides exhibit a more or less abrupt transition from an electrically insulating to a conducting state, and vice versa.\cite{Hickmott1962, Gibbons1964, Dearnale1970, Simmons1971} This phenomenon referred to as resistive switching has experienced a revival in the \red{1990s}, when researchers started thinking about new computer memory concepts. Another remarkable increase of the scientific interest in resistive switching \red{began} in 2008, when Strukov et al. linked resistive switching devices to the memristor postulated by Chua in 1971.\cite{Strukov2008, Chua1971}

During the last 15 years, the switching dynamics of different kinds of complex metal oxides and their potential applications as basis materials for computer memories have been studied.\cite{Beck2000, Seo2004, Rohde2005} The resistive switching random access memory (RRAM) has been identified as a promising candidate for fast terabit memories.\cite{Driscoll2009, Simpson2011, Gerstner2011} Beyond their applications as computer memory, resistive switching devices have also proven to be promising candidates for their applications as lumped elements in neuromorphic systems.\cite{Chang2013, Jeong2013} Here, the key features are low power consumption, passivity, and scalability into the nanometer scale. Also a wide dynamic range is often important in neuromorphic applications, rather than just distinct ``on'' and ``off'' states and ultrafast switching, which are in fact crucial requirements for computer memory applications.

While the technical realization of a resistive switching device can in principle be rather simple -- in its simplest form as a metal-insulator-metal trilayer structure -- the underlying physical mechanisms responsible for the switching are very diverse and complex. A profound understanding is difficult to gain. (For a review of the different physical mechanisms the reader is referred to Waser et al. \cite{Waser2007}). This holds in particular for devices which rely on ionic conduction mechanisms. Here the change in resistance is due to the formation and dissolution of electrically conducting paths in a solid state electrolyte. This phenomenon is also referred to as electrochemical metallization or metallic bridging. On a macroscopic level the mechanism of resistive switching seems to be rather simple: Metal atoms from the active electrode are oxidized and drift/diffuse through the solid state electrolyte towards the opposite (often inert) electrode. Here, they can be reduced and start clustering. Step by step conducting metal filaments grow. However, on the microscopic scale a large number of different physical and chemical reactions might occur. All of them contribute to and guide (to a certain extent) the growth dynamics of the conducting filaments and thus of the global device characteristics. In order to understand the physics behind resistive switching, an elaborate model is needed, which allows for all important physical and chemical mechanisms.

The purpose of this work is -- in contrast to other approaches presented by others so far -- to provide a model which is not limited to i) fast time scale switching dynamics and ii) one growth and dissolution cycle only. In particular, consecutive cycles are simulated. The implemented kinetic Monte-Carlo model consistently coupled with an electric field solver is applied to the formation and dissolution of silver filaments in an Ag/TiO$_2$/Pt sandwich-like thin film system. We find that the numerical results are in excellent agreement with experimentally obtained data. \red{Furthermore}, we observe an unexpected growth mode. In contradiction to the widely acknowledged picture of filament growth from the inert electrode towards the active electrode, we observe filaments that start growing from the active electrode towards the inert electrode.

\section{Simulation Approach}

In order to consistently model the formation and dissolution of conducting filaments we have to cope with a serious time scale problem: An ion performs a ``jump'' from one minimum of the lattice potential to another approx. every 10\textsuperscript{-6}\,s, whereas the actual hopping process takes about 10\textsuperscript{-12}\,s only. Thus, we have to model a so-called infrequent-event system, where the dynamics is characterized by occasional transitions from one state to another, with long periods of relative inactivity between the transitions. This is the reason why the molecular dynamics approach, which has been recently applied by Onofrio et al. \cite{Onofrio2015} to study resistive switching on the ultrafast time scale, is not applicable in our case. The ionic transport which is the basis of the dynamics of the filament growth is much too slow to be described with those kinds of simulation techniques. With respect to rare jumps, the system performs nothing but a simple Markovian walk.\cite{Voter2007} It is rather natural to use the kinetic Monte-Carlo approach which is based on the coarse-graining of the temporal evolution to the discrete rare events.\cite{Fichthorn1990}

The kinetic Monte-Carlo method has been already proven to be suitable for the simulation of filament growth and dissolution in memristive electrochemical metallization devices.\cite{Dirkmann2015} However, only a few results have been published so far to study resistive switching on a microscopic scale.\cite{Pan2011, Jameson2011, Jameson2012, Li2012, Qin2015} All the investigations are mainly focusing on initial filament growth rather than on complete formation and dissolutions cycles. Recently, Menzel et al. reported on results from two-dimensional kinetic Monte-Carlo simulations which allow for several leading physical and chemical processes underlying resistive switching in electrochemical metallization cells.\cite{Menzel2015} They provide a deeper insight into the evolution of the growth and dissolution of metal filaments. However, their results are restricted to the fast (microseconds) time scale.

\begin{figure}[t!]
\centering\includegraphics[width=8cm]{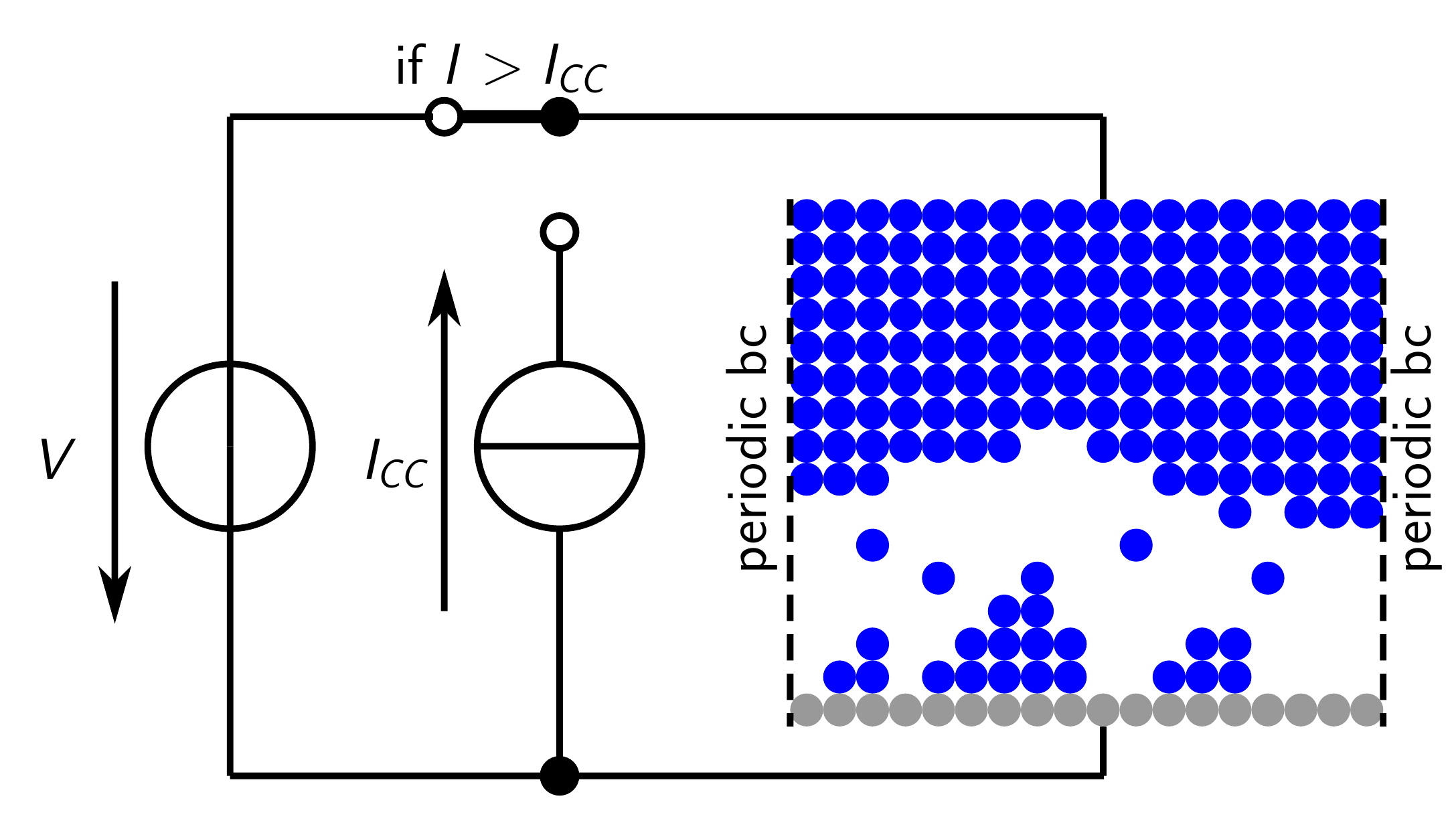}
\caption{Scenario and simulation domain. TiO$_2$ layer (white) between an upper silver electrode (blue dots) and a lower inert platinum electrode (grey dots).}
\end{figure}

Our simulation scenario of a generic electrochemical metallization device is depicted in Fig. 1. It consists of an active metal top electrode (the atoms are represented by blue dots), an inert metallic bottom electrode (gray dots), and a dielectric matrix (white area) which represents the solid state electrolyte between the two electrodes. The growths of conductive metal filaments at the inert metal is based on the transport of metal ions from the active electrode through the solid state electrolyte. The pathway of a metal atom is subject to the electric field due to the external applied voltage $V$. 

\red{It is important to state that in our approach the electric field is \emph{not} calculated from Poisson's equation. This is not feasible since Poisson's equation, derived from Gauss' law $\epsilon_0\nabla\cdot\vec E=\rho$ and $\vec E=-\nabla\Phi$ needs the charge density, which is not known. This approach would be suitable for non-conductive dielectric media and known charge density, but unfavorable if an electric current is flowing. By the way, it is interesting to see that all simulation approaches presented so far -- except for Menzel's approach -- rely on Poisson's equation derived from Gauss' law. One can speculate that this is the reason why only the growth of filaments is simulated but not the transition from a low conductive to a high conductive state, and vice versa. However, this is exactly the situation which we have to cope with. The highly conductive filaments short the low conductive solid state electrolyte and a relatively high electric current is flowing.}

\red{In our approach we solve the continuity equation $\nabla\cdot\vec j=0$ in conjunction with a constitutive law which couples locally the electric field $\vec E$ and the current density $\vec j$. Here, we neglect the displacement current, since it scales with $(L/Tc)^2$ and is therefore very small compared with the conduction current. In this scaling $c$ is the speed of light and $L$ and $T$ are the typical length and time scales of the system.	We apply the manifestation of Ohm's law given by $\vec j(\vec r)=\sigma(\vec r)\vec E(\vec r)$ with $\sigma(\vec r)$ being the conductivity of the respective material, either the solid state electrolyte or the metal.} Since the dynamics of the system can be assumed to be quasi-stationary the electrostatic approximation of Faraday's law, $\nabla\times\vec E=0$, is valid, and therefore $\vec E=-\nabla\Phi$. Combining all equations, we obtain an equation for the quasi-stationary electric potential,
\begin{gather}
\nabla\cdot\left[\sigma(\vec r)\nabla\Phi(\vec r)\right]=0.
\end{gather}
This potential equation is subject to Dirichlet boundary conditions at the top and the bottom boundary of the simulation domain and periodic boundary conditions at the left and the right boundary of the simulation domain. (See Fig. 1). Altogether, the boundary value problem is of elliptic type which we solve numerically using a preconditioned generalized minimal residual scheme.

To be as close as possible to experimental conditions we use an ideal voltage source which is connected to the upper boundary of the top electrode while the bottom electrode is held at $0$ V throughout all simulation cases. The voltage source provides a voltage ramp of 0.5 V/s. The system is driven by this voltage ramp as long as the current $I$ through the device is smaller than or equal to a maximum compliance current $I_{CC}$. In all simulation cases the compliance current is set to $I_{CC}=100$\,$\mu$A. The current $I$ itself is calculated from the local electric field by 
\begin{gather}
\red{I=\int_S \sigma(\vec r)\vec E(\vec r) \cdot \vec n d a.}
\end{gather}
Since we imply $\nabla\cdot\vec j=0$ this integral can be evaluated at an arbitrary vertical position. At the instant of time when $I = I_{CC}$ the voltage source supplying $V(t)$ is substituted by an ideal current source providing the constant compliance current $I_{CC}$. When the value of $V(t) = 0.7$ V is reached, the voltage ramp (i.e., the slope of the source voltage) is reversed to $dV/dt=-0.5$ V/s. (The value of 0.7 V is particularly chosen in order to ensure switching.) Once the actual voltage drop across the device becomes smaller than $V(t)$ (due to the device's negative differential resistance), the current source is re-substituted by the voltage source. The supplied voltage subsequently crosses zero and the current as well as the voltage itself become negative. When it reaches the value of $V(t) = -0.35$ V the ramp changes again to a positive slope and increases until it reaches zero again. Now the final state of the system (i.e., the distribution of atoms/ions) is dumped for consecutive cycles and the simulation stops.

\begin{figure*}[t]
	\centering
	\includegraphics[width=8cm]{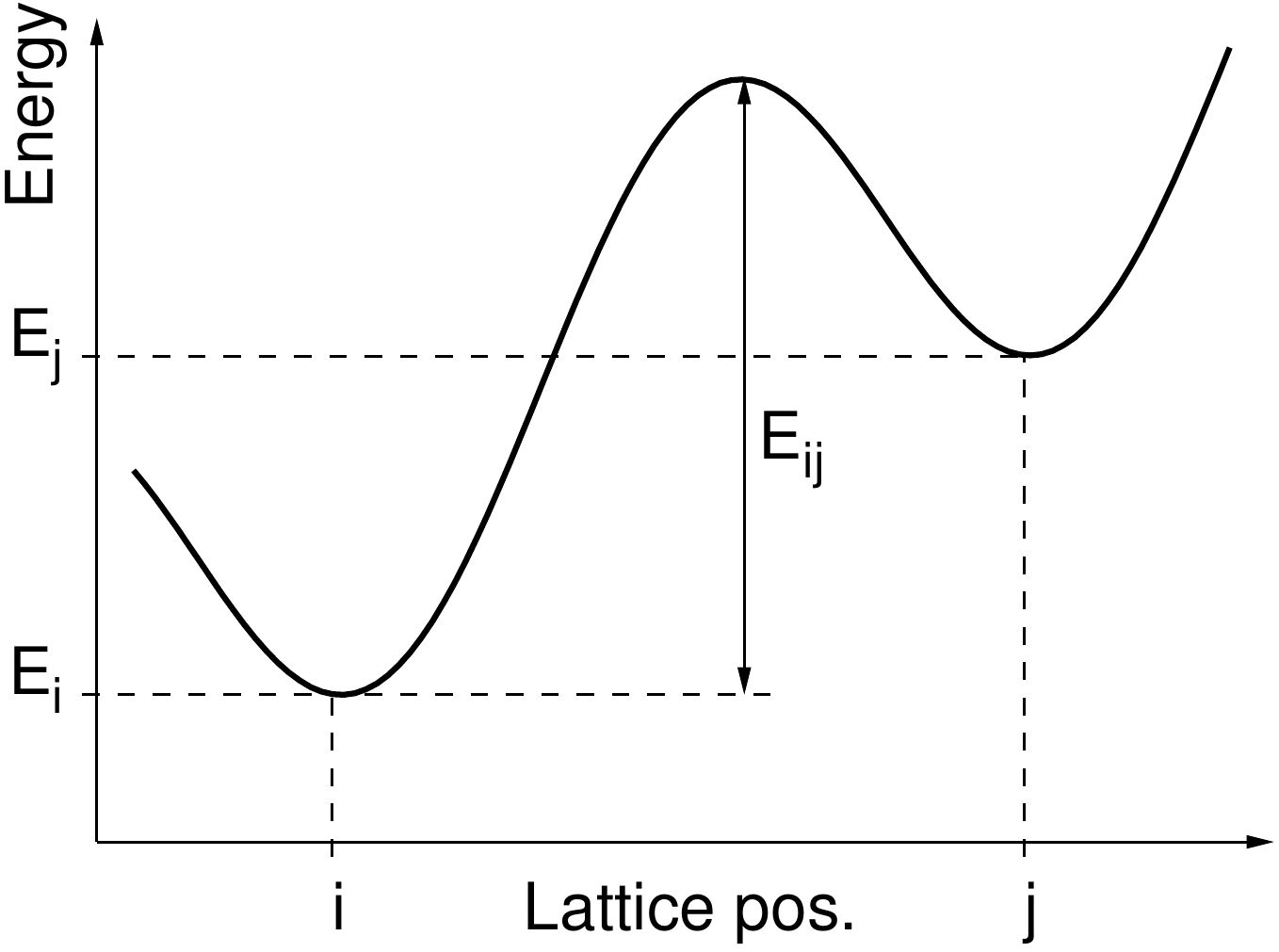}
	\caption{\red{Lattice potential structure for a non-vanishing external electric field.}}
\end{figure*}

In the frame of the kinetic Monte-Carlo approach the ion transport is assumed to be a hopping process on a lattice which represents the solid state electrolyte. The process is driven by the electric field in terms of the potential energy difference between two neighboring lattice sites. The hopping or reaction rates are given by an Arrhenius law according to 
\begin{gather}
k_{ij}=\nu_0\exp{\left[-E_{ij}/k_B T\right]}
\end{gather}
with $\nu_0$ the phonon frequency, $k_B$ the Boltzmann constant, $T$ the lattice temperature, and $E_{ij}$ the energy barrier for the hopping process. \red{The energy barrier $E_{ij}$ can be decomposed into two parts: The first part is the activation energy for the respective process without an external electric field. The second part is the correction of the energy if an external electric field is present. In this case the periodic structure of the lattice potential is bended as it is indicated in Fig. 2. In the frame of the transistion state theory, which is applied here, the correction is a linear interpolation between the two values of the potential at the position $i$ and $j$. Thus, one obtains
\begin{gather}
E_{ij}=E_a+\frac{1}{2}\left(E_j-E_i\right).
\end{gather}
Depending on the spatial position of the metal ion, oxidation at the active electrode, reduction at the active electrode as an adatom, inside a hole or at a kink site, as well as the reduction at the inert electrode may occur and have to be taken into account.\cite{ Pan2011, Jameson2011, Jameson2012, Li2012, Qin2015, Menzel2015} Particularly, the redox reactions at the electrodes are important. The related rates of oxidation of metal atoms and reduction of metal ions are assumed to be constants weighted by Arrhenius-like probabilities which depend on the local electric field and therefore on the local potential. With this approach we are able to mimic redox reactions consistent with the Butler-Volmer theory. It is of course important for the reliability of the numerical results to make sure that all first order processes are incorporated. For our approach we believe that this is given.}

We incorporate the pathway of a metal atom as follows: With a certain probability, depending on the local electric field, a metal atom of the top electrode undergoes oxidation and is released into the solid state electrolyte as a metal ion. Now, it is transported subject to i) the electric field due to the externally applied voltage $V$, ii) the electrical field due to other ions within the solid state electrolyte, and iii) the spatial potential structure of the assumed solid state matrix. When the metal ion arrives at the bottom electrode it meets either an atom of the inert metal and forms a stable nucleus with a certain probability or it meets a previously reduced metal atom from the top electrode. As a third possibility, the ion may not be directly reduced, but diffuses along the electrode surface until it is reduced.
 
\begin{table}[h]
\centering
\begin{tabular}{ll}
\hline\hline
Physical quantity & Value  \\\hline
Temperature &  300 K \\
Phonon frequency   &  1.0$\times$10\textsuperscript{12} Hz   \\
Lattice constant & 3.70$\times$10\textsuperscript{-10} m \\ 
Conductivity of silver & 6.30$\times$10\textsuperscript{7} S/m \\ 
Conductivity of titatium \red{dioxide} & 1.42$\times$10\textsuperscript{2} S/m \\ 
Relative permittivity of titatium \red{dioxide} & 100 As/V \\
$E_a$ for oxidation at silver electrode    &  0.67 eV  \\
$E_a$ for reduction at silver electrode  &  
0.64/0.62/0.6 eV  \\
(adatom/kink/hole)   &   \\
$E_a$ for reduction at platinum electrode   &  0.8 eV\\
\hline\hline
\end{tabular}
\caption{Parameters for all simulation cases}
\end{table} 
  
In our kinetic Monte-Carlo approach the dynamics of the system evolves as follows: For a given spatial distribution of atoms the electric potential -- and also the electric field and the current density -- is calculated subject to the given boundary conditions. A subset of atoms/ions is chosen randomly and individually they undergo randomly chosen processes (oxidation, reduction, or just diffusion). The probability of a process depends on the local electric field. After all chosen atoms/ions have performed their process, their new spatial distribution is recorded and the temporal evolution is realized by enhancing the simulation time $t\rightarrow t+\Delta t$. It is important to note that the time step $\Delta t$ is not prescribed but is calculated by
\begin{gather}
\Delta t=-\frac{\log r}{k_{max}}\quad\mathrm{with}\quad r\in\left[0,1\right].
\end{gather} 
$r$ is a uniformly distributed random number and $k_{max}$ is the maximum hopping rate calculated within this kinetic Monte-Carlo cycle. Inherently, more stable states (indicated by large activation energies) are consistently allowed for by larger time steps. When the time step has been calculated the next cycle starts with the calculation of the electric potential subject to the new spatial distribution of the ions and updated boundary conditions.

\section{Results and Discussion}

\begin{figure*}[t]
\centering
\includegraphics[width=\textwidth]{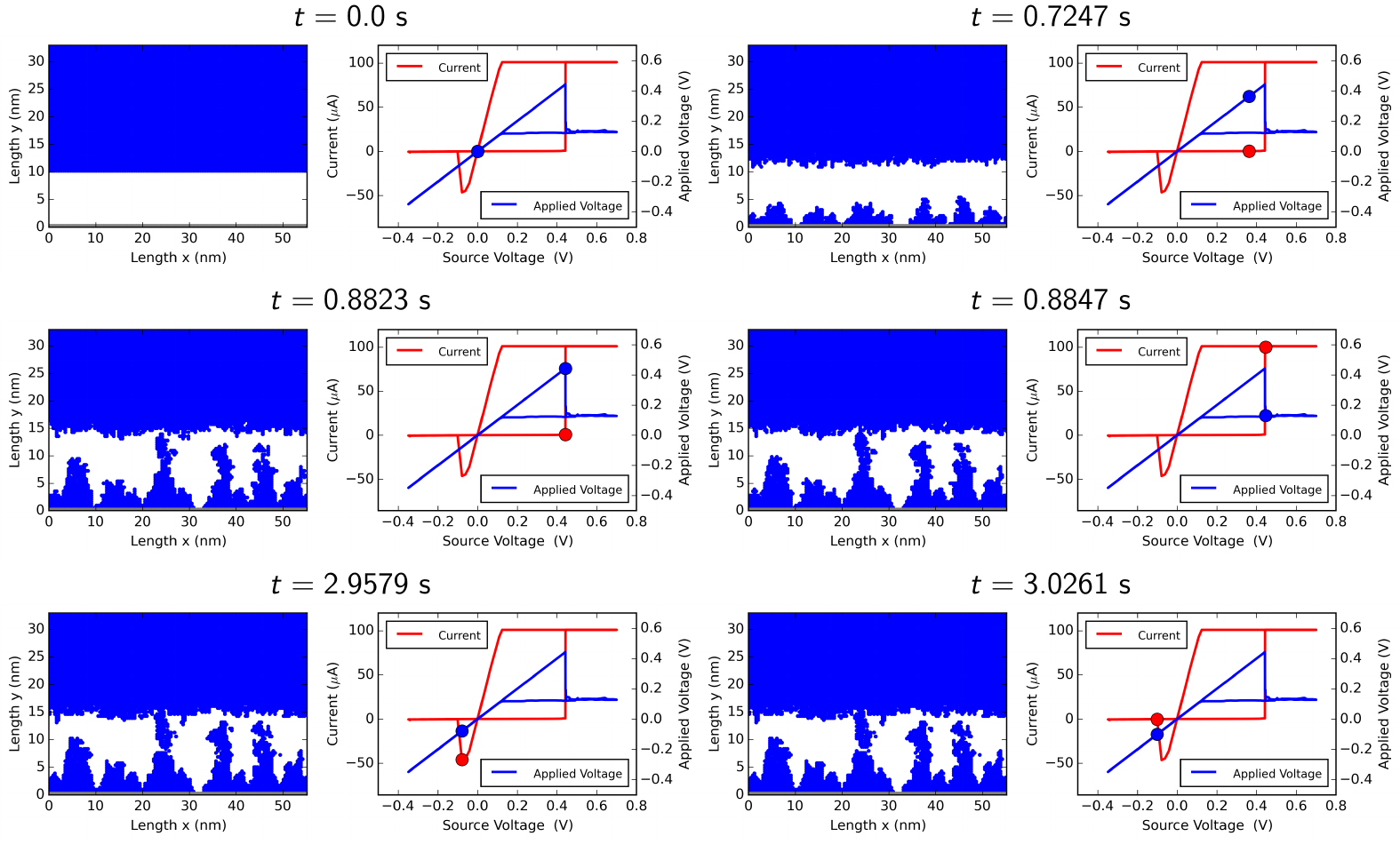}
\caption{Sequence of states at six different instants of time during the electroforming cycle. Status of the filament growth (lhs), the respective electrical state (rhs) marked as a point on the current-voltage curve for the whole cycle (red line), and the voltage at the north electrode as a function of the ramp voltage of the voltage source.}
\end{figure*}

In order to show the performance and the validity of the developed and implemented kinetic Monte-Carlo scheme we investigate the experimentally well-tested Ag/TiO$_2$/Pt system.\cite{Yang2008} The 2D structure under investigation (see Fig. 1) consists of a silver electrode with a thickness of 23.33 nm, an inert electrode of one atomic layer, and a titanium dioxide matrix with a thickness of 10 nm located between the two adjacent electrodes. The length of the sandwich structure is 55.55 nm. Further, a lattice constant of 0.37 nm is used in the simulation which corresponds to titanium \red{dioxide}. All other parameters used in the simulations are listed in Table 1.

The numerical results for an electroforming cycle are provided in Fig. 3. Here, a sequence of different states at six instants of time during the cycle is presented. For each instant of time the status of the filament morphology (lhs) and the respective electrical state (rhs) is depicted. The electrical state is marked by a point on the current-voltage curve for the whole cycle (red line). Additionally, the applied voltage at the top  electrode as a function of the ramp voltage of the voltage source $V(t)$ (blue line) is provided.

The simulation performed for Fig. 3 is a so-called electroforming cycle, i.e., an initial filament growth and dissolution cycle in order to condition the device. We start with the homogeneous situation at $t=0.0$ s (see Fig. 3, upper left). The inert platinum (bottom) electrode is indicated by one thin monolayer of individual (gray) atoms whereas the silver (top) electrode is constituted of 9,450 individual (blue) atoms. The source voltage applied to the upper layer of the top electrode is initially $V=0.0$ V. Consequently, the current through the device is $I=0.0$ A. With increasing time, the bias voltage increases according to the defined ramp of 0.5 V/s. With increasing voltage the probability for Ag atoms of the top electrode to oxidize also increases. Now, that they are positively charged ions after oxidation, they are released into the solid state electrolyte and drift due to the electric field towards the Pt (bottom) electrode. This drift is predominantly due to the external electric field provided by the applied voltage. A minor contribution stems from all other positively charged ions which are present elsewhere in the solid state electrolyte. When the ions reach the bottom electrode (constituted of either Pt atoms or earlier arrived Ag atoms) they might be reduced, become neutral atoms, stick to the surface, and start to cluster. Of course reduced Ag atoms can in principle re-oxidize again. This is a process which is incorporated. \red{However, at this stage of the dynamics the probability for this process is rather small, but not zero, and only rarely observed.}

One can observe that the growth of Ag filaments is very inhomogeneous (see Fig. 3, e.g. upper right). This can be understood as follows: In our approach the nucleation of Ag occurs at random positions at the Pt electrode. We do not set an artificial nucleation seed as it has been done by others ``in order to save computational time''.\cite{Menzel2015} At these positions Ag atoms tend to cluster. This tendency is even amplified by the fact that the electric field is enhanced by sharp edges. The ions released from the top electrode are attracted by enhanced electric fields so that they reduce preferably at these very positions. This self-amplifying mechanism turns out to be the main reason for the formation of metal filaments.

\begin{figure}[t]
\centering
\includegraphics[width=8cm]{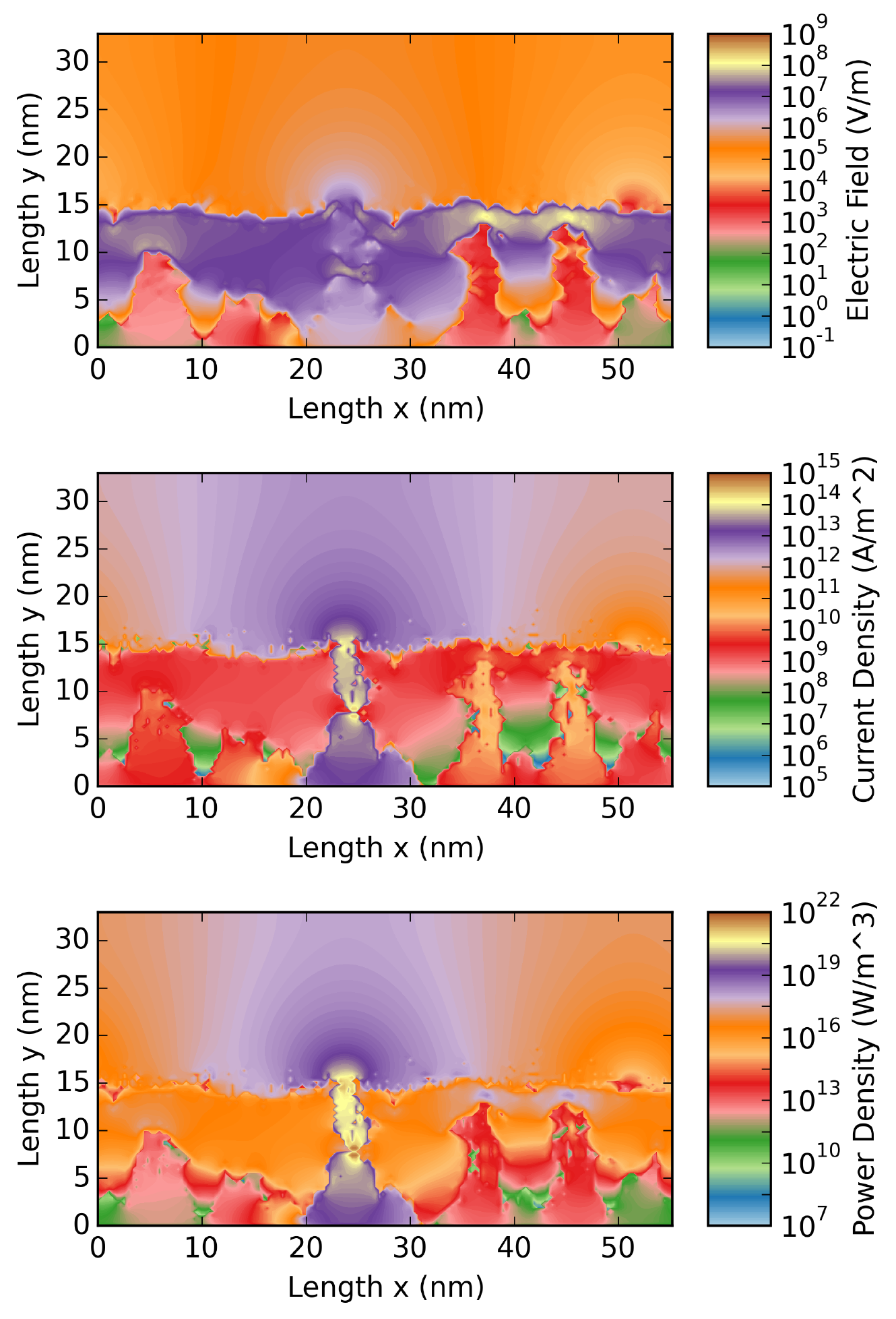}
\caption{Electric field (top), current density (middle), and power density (bottom) at the instant of time $t=2.279$ s which is close to the time of dissolution.}
\end{figure}

With every Ag atom which sticks to the surface the bottom electrode is inhomogeneously growing and the resistance of the device is decreasing. Still, the current is quite small since the low-conductive solid state electrolyte determines the resistance of the device. The growth process continues until the current $I$ reaches the prescribed compliance current $I_{CC}$. From now on the current is kept constant. Of course the filaments still grow and the resistance still drops. This leads to a decrease of the actual voltage at the top electrode (blue lines in Fig. 3.), which continues until the growth of the filament stops and the resistance saturates. 

As described above, the current source is again re-substituted by the voltage source at a certain instant of time. The voltage ramp is now negative. The voltage crosses zero and becomes increasingly negative. As a result the current is also negative and increases in magnitude. When the current becomes large enough the filaments start to dissolve. The device switches back to the low conductive state. The voltage ramp becomes positive again. When the voltage reaches zero the simulation stops. 

Of particular importance is the onset of the filament dissolution. In order to discuss this mechanism in more detail the electrical field strength at the time of the onset of the dissolution ($t=2.279$ s) is calculated and shown in Fig. 4 (top), together with the current density (Fig. 4, middle), and the power density (Fig. 4, bottom) calculated by $\vec E\cdot \vec j$. In the highly conductive regions of the metal filaments the electric field is small whereas the current density is very high. The power density is very high only at certain positions inside the filaments. One can speculate that thermal effects are very strong at these positions. Also the effect of electromigration should be significant here. In the simulations we localize the region of highest power density and initiate the dissolution of the filament at this very position. The probability for this process is assumed to be depending on the local current density. 

Going back to Fig. 3, one can observe that the reset is a stepwise process. This is consistent with experimental investigations and recently reported simulation results on a similar system.\cite{Russo2009,Menzel2015} One can also find, that the overall switching dynamics of the electroforming cycle is close to realistic findings. Once electroformation is finished, it is interesting to notice that the low conductive state is realized by the absence of a few atoms only. This can be seen in the lower right of Fig. 3.

\begin{figure*}[t]
\centering
\includegraphics[width=\textwidth]{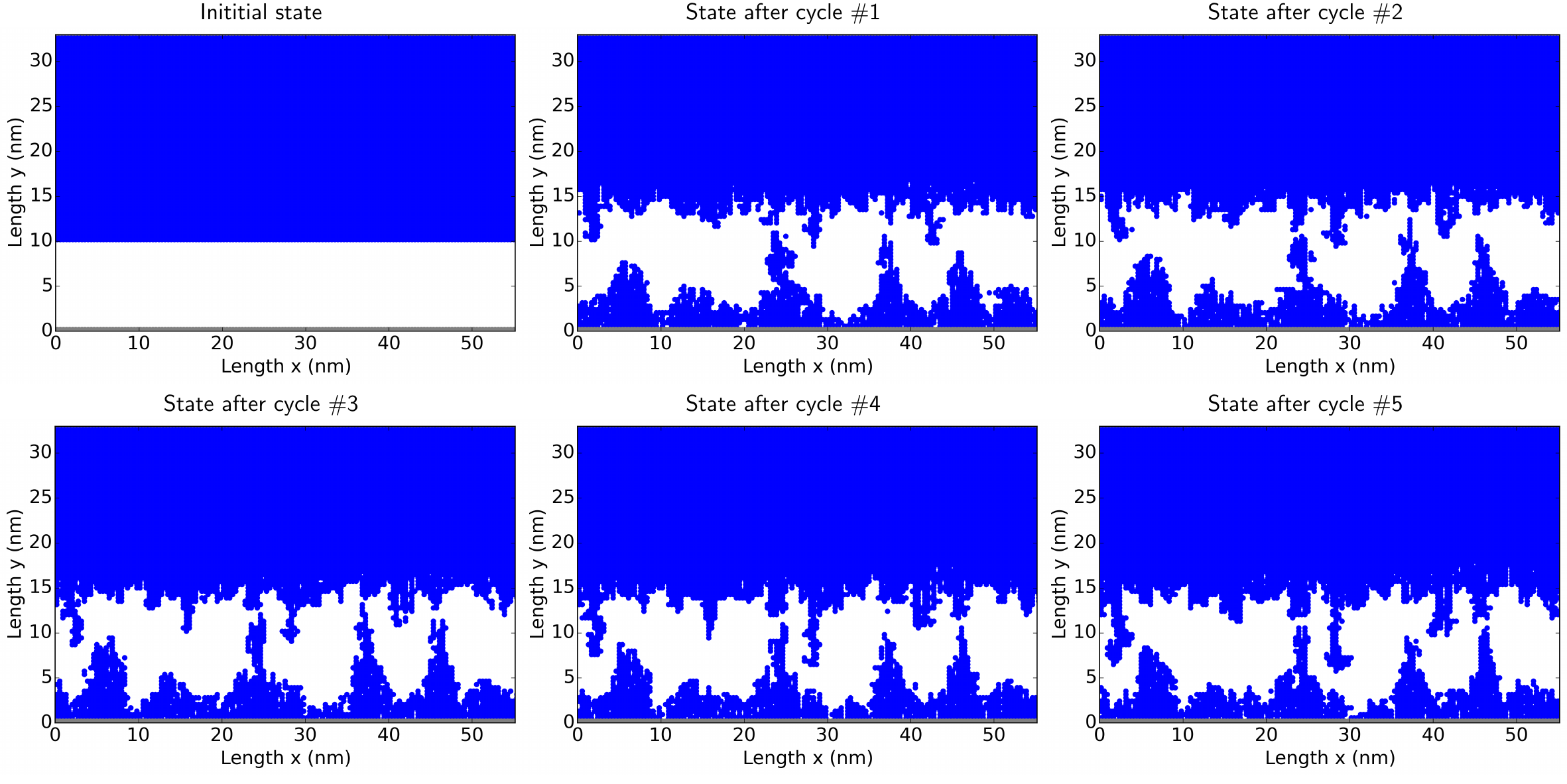}
\caption{Initial and final states of five consecutive set and reset cycles.}
\end{figure*}

In this work we are interested in the dynamics of filament growth and dissolution in some more detail. Therefore, we do not restrict ourselves to the electroforming cycle. We also simulate a number of consecutive set and reset cycles. The final state of a complete cycle is used as the initial state of the subsequent cycle. These individual final states are shown in Fig. 5 (together with the homogeneous initial state). It can be observed that the final states do not differ much. Interestingly, the overall morphology of the filaments is preserved after every set-reset cycle, even if the device is in a low conductive state. In the low conductive state the filaments are by far not dissolved completely. Moreover, we found that the switching from the high conductive to the low conductive state, and vice versa, also for consecutive cycles is realized by local oxidation and reduction of a few atoms only. 

\begin{figure}[t]
\centering
\includegraphics[width=8cm]{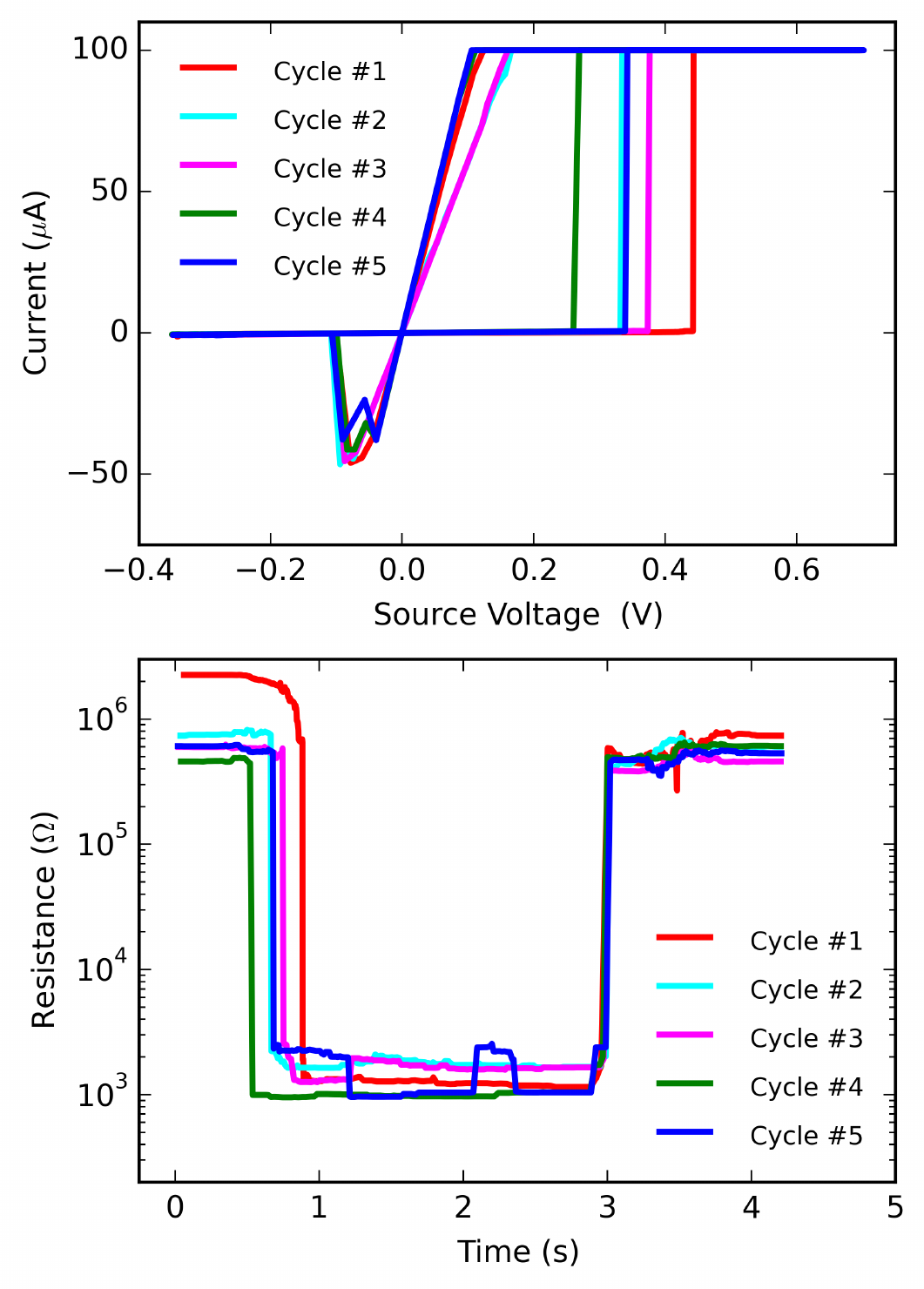}
\caption{Current-voltage characteristics (top) and resistances as functions of time (bottom) for five consecutive cycles.}
\end{figure}

It is important to note that the microscopic structure does not influence the global device characteristic with respect to the resistive behavior. This can be seen in Fig. 6 where the current voltage characteristics (Fig. 6, top) and the resistance as a function of time (Fig. 6, bottom) are shown for the five consecutive cycles of Fig. 5. One can clearly observe that the electroforming cycle takes a longer time than all consecutive cycles. Consequently, the switching voltage from low-conductive to high-conductive state is larger. The reason is of course that a large number of Ag atoms has to be transported first in order to build an initial filament. The opposite holds for all consecutive cycles. Moreover, it is interesting to note that a larger reset voltage is needed to dissolve the filament. For all consecutive cycles the dynamics is similar: The set and reset voltages for the different cycles are almost the same and the low and the high resistance states are stable during the cycles. Hence, reliably distinct states can be realized and predicted. This is in agreement with experimental findings.\cite{Li2011}

In Fig. 5 one can observe another very interesting and potentially important phenomenon. One can find an additional filament growth mode which starts from the active top electrode towards the inert bottom electrode. This result contradicts the widely acknowledged picture of filament growth only from the inert electrode towards the active electrode. However, such an unexpected growth mode has also been experimentally observed.\cite{Yang2012} It has been worked out recently by Yang et al. that such a specific growth mode functionally relies on the particular microstructure and kinetic factors of the material system.\cite{ Yang2014} By investigating metal clusters inside a dielectric layer they have argued that these metal clusters can be treated as bipolar electrodes influencing the growth mode significantly. Moreover, they have shown evidence that the filament growth is guided by mainly two kinetic factors, i.e. ion mobility and redox rates, both of which are incorporated in the our kinetic Monte-Carlo model.\cite{Yang2014} Therefore, we strongly support the experimental findings of Yang et al. 

\section{Conclusions}

We report on kinetic Monte-Carlo calculations of resistive switching and the underlying growth dynamics of conductive filaments in electrochemical metalization devices consisting of a Ag/TiO$_2$/Pt sandwich-like thin film system. The developed model is not limited to i) fast time scale dynamics and ii) only one growth and dissolution cycle of metallic filaments. In particular, the presented results from the simulation of consecutive cycles are in excellent agreement with experimentally obtained data. Most importantly, we also find an unexpected filament growth mode which is in contradiction to the widely acknowledged picture of filament growth from the inert electrode towards the active electrode but consistent with recently published experimental findings. We observe filaments that start growing from the active electrode towards the inert electrode. 

Although the phenomenon of filament growth is inherently three dimensional, we believe that with our model we provide a powerful simulation tool which can be used to gain deeper understanding of the relevant physical and chemical processes for resistive switching on both, atomistic length scales and experimental time scales.

\section*{Acknowledgments}
The authors gratefully acknowledge financial support by the Deutsche Forschungsgemeinschaft in the frame of Research Group FOR 2093 ``Memristive Devices for Neural Systems''. MZ and TM thanks Dr. Doo Seok Jeong (Korea Institute of Science and Technology) for valuable discussions.






\begin{thebibliography}{}
	
\bibitem{Hickmott1962}
T.W. Hickmott, 
J. Appl. Phys. \textbf{33}, 2669 (1962)

\bibitem{Gibbons1964}
J.F. Gibbons and W.E. Beadle,
Solid State Electron. \textbf{7}, 785 (1964)

\bibitem{Dearnale1970}
G. Dearnale, A.M.  Stoneham, and D.V. Morgan,
Rep. Progr. Phys. \textbf{33}, 1129 (1970)

\bibitem{Simmons1971}
J.G. Simmons,
J. Phys. D: Appl. Phys. \textbf{4}, 613 (1971)

\bibitem{Strukov2008}
D.B. Strukov, G.S. Snider, D.R. Stewart, and R.S. Williams,
Nature \textbf{453}, 80 (2008)
	
\bibitem{Chua1971}
L. Chua, 
IEEE Trans. Circuit Theory \textbf{18}, 507 (1971)

\bibitem{Beck2000}
A. Beck, J.G. Bednorz, C. Gerber, C. Rossel, and D. Widmer,
Appl. Phys. Lett., \textbf{77}, 139 (2000)

\bibitem{Seo2004}
S. Seo, M.J. Lee, D.H. Seo, E.J. Jeoung, D.S. Suh,
Y.S. Joung, I.K. Yoo, I.R. Hwang, S.H. Kim, I.S. Byun, 
J.S. Kim, J.S. Choi, and B.H. Park,
Appl. Phys. Lett. \textbf{85}, 5655 (2004)
	
\bibitem{Rohde2005}    
C. Rohde, B.J. Choi, D.S. Jeong, S. Choi, 
J.S. Zhao, and C.S. Hwang,
Appl. Phys. Lett. \textbf{86}, 262907 (2005)
	
\bibitem{Driscoll2009}
T. Driscoll, H.T. Kim, B.G. Chae, M. Di Ventra, and D.N. Basov
Appl. Phys. Lett. \textbf{95}, 043503 (2009)
	
\bibitem{Simpson2011}
R.E. Simpson, P. Fons, A.V. Kolobov, T. Fukaya, 
M. Krbal, T. Yagi, J. Tominaga,
Nat. Nanotechnol. \textbf{6}, 501 (2011)
	
\bibitem{Gerstner2011}
E. Gerstner, 
Nat. Phys. \textbf{7}, 837 (2011)

\bibitem{Chang2013}
T. Chang, Y. Yang, W. Lu,
IEEE Circuits Syst. Mag. \textbf{13}, 56 (2013)
		
\bibitem{Jeong2013}
D.S. Jeong, I. Kim, M. Ziegler, H. Kohlstedt,
RSC Adv. \textbf{3} 3169 (2013)

\bibitem{Zahari2015}
F. Zahari, M. Hansen, T. Mussenbrock, M. Ziegler, and H. Kohlstedt,
AIMS Materials Science, AIMS Materials Science \textbf{2}, 203 (2015)
	
		
\bibitem{Waser2007}
R. Waser and M. Aono,
Nature Mater. \textbf{6}, 833 (2007)
	
					
\bibitem{Onofrio2015}
N. Onofrio, D. Guzman, and A. Strachan,
Nat. Materials \textbf{14}, 440 (2015)
	
\bibitem{Voter2007}
A. F. Voter, \textit{ 
in Radiation Effects in Solids}, edited 
by K. E. Sickafus and E. A. Kotomin (Springer, 2007).

\bibitem{Fichthorn1990} 
K.A. Fichthorn and W.H. Weinberg, J. Chem. Phys. \textbf{95} 1090 (1991)

\bibitem{Dirkmann2015}
S. Dirkmann, J. Trieschmann, M. Hansen, M. Ziegler,
H. Kohlstedt, and T. Mussenbrock, 
Proceeding of Spring Meeting of the German Physical Society DS 38.13, Berlin, Germany (2015) 
	
\bibitem{Pan2011}
F.Pan, S. Yin, and V. Subramanian,
IEEE Electr. Device. L. \textbf{32}, 7 (2011)
	
\bibitem{Jameson2011}   
J. R. Jameson, N. Gilbert, F. Koushan, J. Saenz, 
J. Wang, S. Hollmer, and M.N. Kozicki,
Appl. Phys. Lett. \textbf{99}, 063506 (2011)
	
\bibitem{Jameson2012}
J. R. Jameson, N. Gilbert, F. Koushan, J. Saenz, 
J. Wang, S. Hollmer, and M.N. Kozicki,
Appl. Phys. Lett. \textbf{100}, 023505 (2012)
	
\bibitem{Li2012}
D. Li, M. Li, F. Zahid, J. Wang, and H. Guo,
J. Appl. Phys. \textbf{112}, 7 (2012)

\bibitem{Qin2015}
S. Qin, Z. Liu, G. Zhang, J. Zhang, Y. Sun, H. Wu, H. 
Qian, and Z. Yu,
Phys Chem Chem Phys. \textbf{17}, 8627 (2015)
				
\bibitem{Menzel2015} S. Menzel, P. Kaupmann, and R. Waser,
Nanoscale \textbf{7}, 12673 (2015)

\bibitem{Yang2008}
J.J. Yang, M.D. Pickett, X. Li, D.A.A. Ohlberg, D.R. Stewart, and R.S. Williams, Nature Nanotechnology 3, \textbf{429} (2008) 

\bibitem{Russo2009}
U. Russo, D. Kamalanathan, D. Ielmini, A.L. Lacaita, and M.N. Kozicki, IEEE Trans. Electron Devices \textbf{56}, 1040 (2009)

\bibitem{Li2011} 
H. Li, Y. Xia, H. Xu, L. Liu, X.Li, Z. Tang, X. Chen, A. Li, J. Yin, and Zhiguo Liu
AIP Advances \textbf{1}, 032141 (2011) 
\bibitem{Yang2012} 
Y. Yang, P. Gao, S. Gaba, T. Chang,	X. Pan, and W.D. Lu
Nat. Commun. \textbf{3}, 732 (2012)

\bibitem{Yang2014}
Y. Yang, P. Gao, L. Li,	X. Pan,	S. Tappertzhofen, S.H. Choi, R. Waser, I. Valov, and W.D. Lu
Nat. Commun \textbf{5}, 4232 (2014) 

\end{thebibliography}
\end{document}